\documentclass[9pt, conference]{IEEEtran}
\IEEEoverridecommandlockouts

\usepackage[utf8]{inputenc}

\usepackage[nocompress]{cite}
\usepackage{amsmath,amssymb,amsfonts}
\usepackage{algorithmic}
\usepackage{graphicx}
\usepackage{textcomp}
\usepackage{xcolor}
\usepackage[hyphens]{url}
\usepackage{multirow}
\usepackage{enumitem}
\usepackage{booktabs}
\usepackage{pifont}
\usepackage{color}
\usepackage{tikz}
\usepackage{multirow}
\usepackage{float}
\usepackage{listings}
\usepackage{soul}
\usepackage{fontawesome}
\usepackage{subfig}
\usepackage{textcomp}
\usepackage{scalerel}
\usepackage[hidelinks]{hyperref}
\usepackage{makecell}
\usepackage{cuted}

\usepackage{caption}
\usepackage{booktabs}
\usepackage{enumitem}% http://ctan.org/pkg/enumitem

\let\subparagraph\paragraph % avoid a warning
\usepackage{titlesec}

\titlespacing{\subsubsection}
  {0pt}
  {1ex plus 0.3ex minus 0.1ex}
  {0.5ex plus 0.1ex minus 0.1ex}

\graphicspath{{./figures/}}

\newcommand*\circled[1]{
    \raisebox{.4pt}{
        \tikz[baseline=(char.base)]{
            \node[shape=circle,draw,inner sep=1pt, style={fill=black, text=white}, scale=0.75] (char) {\textbf{#1}};
        }
    }
}

\newcommand{\nth}{$n$\textsuperscript{th}}
\newcommand{\first}{1\textsuperscript{st}}

\newcommand{\arrstream}{\texttt{array\_stream}}
\newcommand{\arrstreams}{\texttt{array\_stream}s}
\newcommand{\cpystream}{\texttt{copy\_stream}}

\newcommand{\ourwork}{INR-Arch}

\begin{document}

\title{\huge INR-Arch: A Dataflow Architecture and Compiler for Arbitrary-Order Gradient Computations in Implicit Neural Representation Processing
% \vspace{-1em}
}

% \and
% \and
% \and
% \and
% \and

\author{
Stefan Abi-Karam*\textsuperscript{1,2},
Rishov Sarkar*\textsuperscript{1},
Dejia Xu\textsuperscript{3}, Zhiwen Fan\textsuperscript{3},
Zhangyang Wang\textsuperscript{3},
Cong Hao\textsuperscript{1}
\\
Georgia Institute of Technology\textsuperscript{1}, Georgia Tech Research Institute\textsuperscript{2}, University of Texas at Austin\textsuperscript{3}
\\
\{%
\href{mailto:stefanabikaram@gatech.edu}{\nolinkurl{stefanabikaram}},
\href{mailto:rishov.sarkar@gatech.edu}{\nolinkurl{rishov.sarkar}},
\href{mailto:callie.hao@gatech.edu}{\nolinkurl{callie.hao}}%
\}@gatech.edu,
\{%
\href{mailto:dejia@utexas.edu}{\nolinkurl{dejia}},
\href{mailto:zhiwenfan@utexas.edu}{\nolinkurl{zhiwenfan}},
\href{mailto:atlaswang@utexas.edu}{\nolinkurl{atlaswang}}%
\}@utexas.edu
}

\maketitle

\thispagestyle{plain}
\pagestyle{plain}

\footnotetext{*Equal contribution.}

\begin{abstract}
An increasing number of researchers are finding use for \nth-order gradient computations for a wide variety of applications, including graphics, meta-learning (MAML), scientific computing, and most recently, implicit neural representations (INRs). Recent work shows that the gradient of an INR can be used to edit the data it represents directly without needing to convert it back to a discrete representation. However, given a function represented as a computation graph, traditional architectures face challenges in efficiently computing its \nth-order gradient due to the higher demand for computing power and higher complexity in data movement. This makes it a promising target for FPGA acceleration.
In this work, we introduce \ourwork, a framework that transforms the computation graph of an \nth-order gradient into a hardware-optimized dataflow architecture. We address this problem in two phases. First, we design a dataflow architecture that uses FIFO streams and an optimized computation kernel library, ensuring high memory efficiency and parallel computation. Second, we propose a compiler that extracts and optimizes computation graphs, automatically configures hardware parameters such as latency and stream depths to optimize throughput, while ensuring deadlock-free operation, and outputs High-Level Synthesis (HLS) code for FPGA implementation. We utilize INR editing as our benchmark, presenting results that demonstrate 1.8--4.8\texttimes{} and 1.5--3.6\texttimes{} speedup compared to CPU and GPU baselines respectively. Furthermore, we obtain 3.1--8.9\texttimes{} and 1.7--4.3\texttimes{} lower memory usage, and 1.7--11.3\texttimes{} and 5.5--32.8\texttimes{} lower energy-delay product. Our framework will be made open-source and available on GitHub.\footnote[7]{\url{https://github.com/sharc-lab/inr-arch}}
\end{abstract}

\section{Introduction}

Implicit neural representations (INRs) are enjoying great popularity for a variety of use cases, including 3D neural rendering and stylization in augmented and virtual reality (AR/VR)\cite{mildenhall_nerf_2020, fan_unified_2022}, application-agnostic data representation and compression\cite{siren, coin,coin++}, and super-resolution and inpainting for data across various modalities, such as images and videos~\cite{siren, chen_videoinr_2022}. A core strength of INRs lies in their capacity for data compression. As an effective, high-fidelity encoding approach for diverse data types, INRs represent a promising path to efficient data management~\cite{coin,coin++}.

However, it is critical to recognize the implications of such compact data encoding for computational hardware requirements. Considering the emerging paradigm where memory costs more than computation, a memory-efficient solution can provide high energy and area efficiency. Therefore, it is crucial to develop methods to perform rapid gradient computations without relying on large, memory-intensive hardware.

Meanwhile, hardware designers are embracing dataflow architectures to achieve low latency through overlapping computation kernels within their designs. Streaming in dataflow designs allows for individual processes to work at a much finer granularity than input and output arrays; instead, they can incrementally produce partial outputs or consume partial inputs through first-in-first-out (FIFO) streams of data. When large numbers of these processes are combined in this way, massive latency savings can be achieved effectively by exploiting the throughput of each process.

Motivated by the need for efficient INR computation and editing, we propose \ourwork{} with emphasis on a dataflow architecture and specialized compiler. Our key contributions are as follows:

\begin{itemize}
    \item[\circled{1}] \textbf{Dataflow Architecture}: We propose a dataflow architecture based on FIFO-based array streams and a library of optimized computation kernels that operate on array streams. This approach allows for increased memory efficacy and overlapping computations.
    \item[\circled{2}] \textbf{Computation Graph Extraction \& Optimization}: We propose an automated method to extract the computation graph of the gradient of a PyTorch tensor, along with several lossless optimization techniques to simplify the resulting graph.
    \item[\circled{3}] \textbf{Deadlock Analysis and Optimization}: We propose a novel technique to quickly and accurately determine whether a given set of FIFO depths will cause a dataflow design to deadlock.
    \item[\circled{4}] \textbf{FIFO Depth Analysis and Optimization}: We extend the deadlock analysis to compute latency estimates based on a set of FIFO depths, and we propose a procedure to quickly determine a reduced set of FIFO depths that lowers memory usage without impacting performance.
    \item[\circled{5}] \textbf{Code Generation}: We propose a compiler that uses the processed computation graph and a set of FIFO depths to generate a dataflow architecture that executes the graph in hardware.
    \item[\circled{6}] \textbf{Power, Latency, and Memory Improvements vs. CPU \& GPU}: We evaluate \ourwork{} applied to INR editing targeting an FPGA platform and compare its latency, memory usage, and energy-delay product against CPU and GPU baselines.
\end{itemize}

\section{Background and Motivation}

\subsection{High-Order Gradients}
\label{sec:background-high-order-gradients}

Popular machine learning frameworks such as PyTorch and TensorFlow are capable of automatically computing arbitrary-order gradients of a given function through reverse mode automatic differentiation~\cite{baydin2018automatic}. This involves first representing the function as a computation graph, where each node represents a primitive operation such as elementwise add, transpose, or matrix multiply. The framework then recursively applies the chain rule of differentiation on the graph to obtain a new computation graph, which represents a function whose output is the gradient of the original function. This automatic differentiation process can be repeated recursively to obtain second- and higher-order gradients of a function.

Higher-order gradients play a crucial role in various fields, such as scientific computing, computer graphics, and deep learning. For instance, in scientific computing, higher-order gradients are essential for accurately modeling complex problems in areas such as fluid dynamics \cite{anderson_computational_1995}. 

Similarly, in computer graphics, higher-order gradients are traditionally used to render images with high fidelity and realism. More recently, differentiable rendering techniques \cite{kato_differentiable_2020} have been developed to work along side deep learning to incorporate graphics rendering in the end-to-end training pipeline. Neural radiance fields (NeRF) \cite{mildenhall_nerf_2020} are a popular example of differentiable rending for deep learning.

Moreover, the use of higher-order gradients is becoming increasingly popular in the field of deep learning. Higher-order gradients are used in the traditional training of models as well as for meta-learning such as model-agnostic meta-learning (MAML) \cite{finn_model-agnostic_2017} and hyperparameter optimization \cite{maclaurin_gradient-based_2015}. Additionally, higher-order gradients have been shown to be an effective tool for processing implicit neural representations (INRs) to apply arbitrarily learnable data transformations efficiently.

\subsection{Implicit Neural Representations}

\begin{figure}
    \centering
    \includegraphics[width=1.0\linewidth]{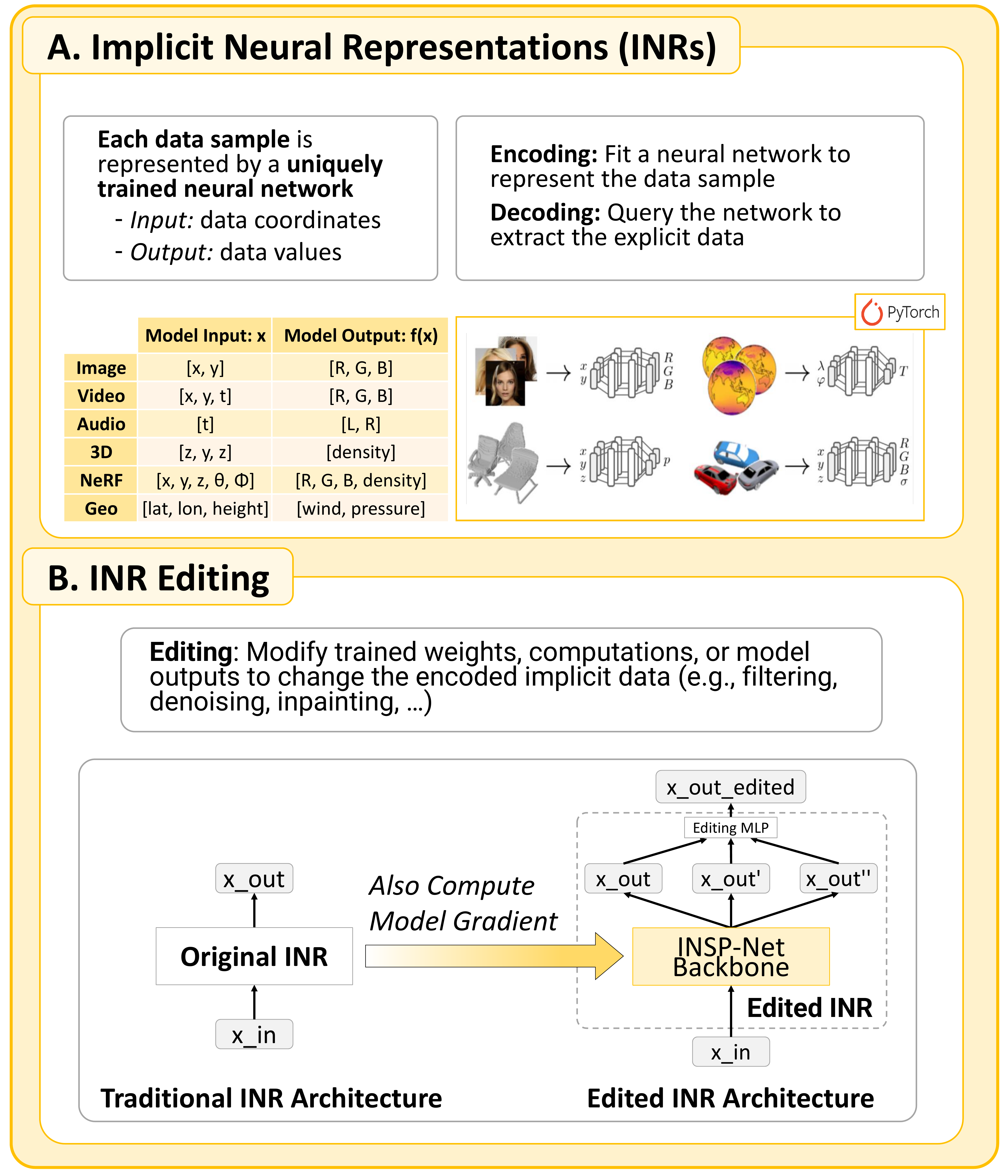}
    \caption{A visual overview of \textbf{A)} Implicit Neural Representations (INRs), and \textbf{B)} INR Editing using the INSP-Net architecture\cite{inrdsp}.}
    \label{fig:inr_fig}
\end{figure}

Implicit neural representations (INRs) are a way to represent individual data points as entire neural networks, as shown in Fig.~\ref{fig:inr_fig}A. Given a single data sample, such as an image, audio file, 3-D model, etc., and a suitable neural network architecture~\cite{siren}, the data sample can be \textit{encoded} as a set of weights and biases for the neural network and later \textit{decoded}, i.e., reconstructed, from the weights.

Encoding an INR involves training the chosen neural network architecture to predict output coordinates from input coordinates within a single data sample, effectively overfitting the neural network to this one sample. For instance, to encode an image file, we first consider the image to be a ``training dataset'' for the neural network consisting of input $(x, y)$ 2-D coordinate pairs mapped to 3-D outputs, representing the red, green, and blue (RGB) colors of the pixel at the input $(x, y)$ coordinates within the image. After training the neural network to predict these mappings, the neural network weights and biases can themselves be considered an \textit{implicit} representation of the image in \textit{weight-space}, i.e., an INR.

It follows that decoding an INR involves plugging in discrete input coordinates into a neural network with weights and biases given by the INR to obtain output coordinates; in the case of images, this means plugging in $(x, y)$ coordinate values and obtaining RGB color values in \textit{pixel-space}.

INRs are useful for several purposes. \underline{First,} since the inputs and outputs of the INR are continuous coordinate values, the INR can be treated as a continuous representation of a discrete input sample, allowing for super-resolution beyond that of the input sample. For instance, given an image encoded as an INR, during decoding, we can plug in non-pixel-aligned input coordinates and obtain output colors corresponding to points between pixels of the original image, effectively providing unlimited image resolution. \underline{Second,} INRs can be used as an effective compression scheme, as the weights in an INR can require less space than the original data while still maintaining high fidelity~\cite{dataasfuncta,coin,coin++}. \underline{Third,} INRs are a universal representation of any type of data that can be represented as a mapping from input coordinates to output coordinates, making them versatile for compression and super-resolution for a wide variety of data formats.

\subsection{INR Editing}

A recent work by Xu \textit{et al.}~\cite{inrdsp} demonstrates that for images encoded as INRs, we can operate directly on the \textit{weight-space representation} of the image to obtain another INR, using a different neural network architecture dubbed \textbf{INSP-Net} (shown in Fig.~\ref{fig:inr_fig}B), whose decoding corresponds to a desired signal processing transformation of the original image in \textit{pixel-space}, such as blurring, de-noising, etc. In other words, if we want to edit an image encoded as an INR by, e.g., blurring it, we do not have to decode the INR to pixel-space, apply a blur filter, then re-encode to another INR. By computing the model output and up to the \nth-order gradients of the output as input features for a trainable MLP, specific signal processing tasks can be achieved on a distribution of data. However, computing higher-order gradients needed for editing is complex, resulting in exponentially more complex computation graphs as the number of gradients increases. This provides a key motivation
for hardware acceleration of the exact gradient computation.

\section{Proposed Methodology}
\begin{figure*}[!ht]
    \centering
    \includegraphics[width=\textwidth]{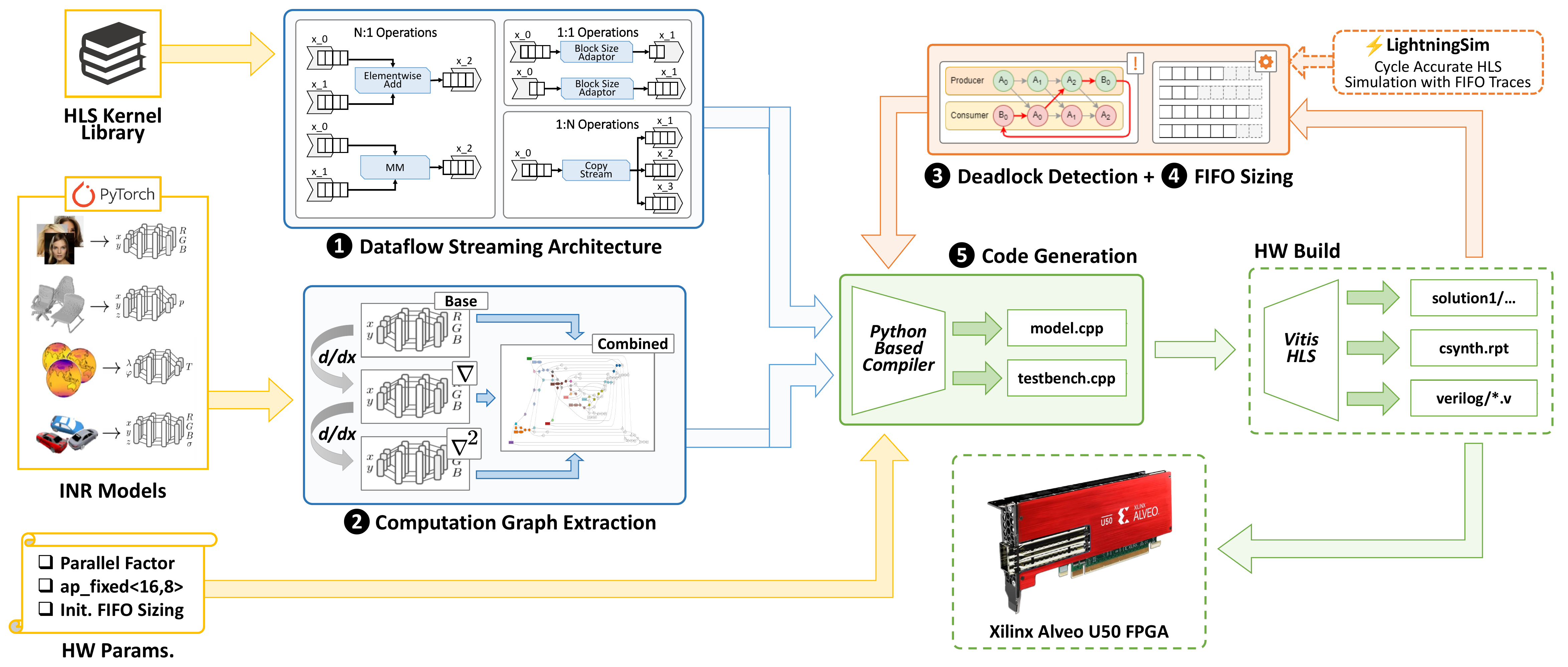}
    \caption{An overview of the \ourwork{} framework for end-to-end hardware acceleration for INR editing based on the INSP-Net~\cite{inrdsp} architecture.}
    \label{fig:main_fig}
\end{figure*}

With this motivation in mind, we propose \textbf{\ourwork{}}, a dataflow architecture and compiler for arbitrary-order gradient computations in INR processing. We first discuss the \ourwork{} dataflow architecture, followed by our compiler flow that translates gradient computations in PyTorch to a synthesizable and performant HLS design.

\subsection{Dataflow Architecture}
\label{sec:dataflow-arch}

\begin{figure}
    \centering
    \includegraphics[width=\linewidth]{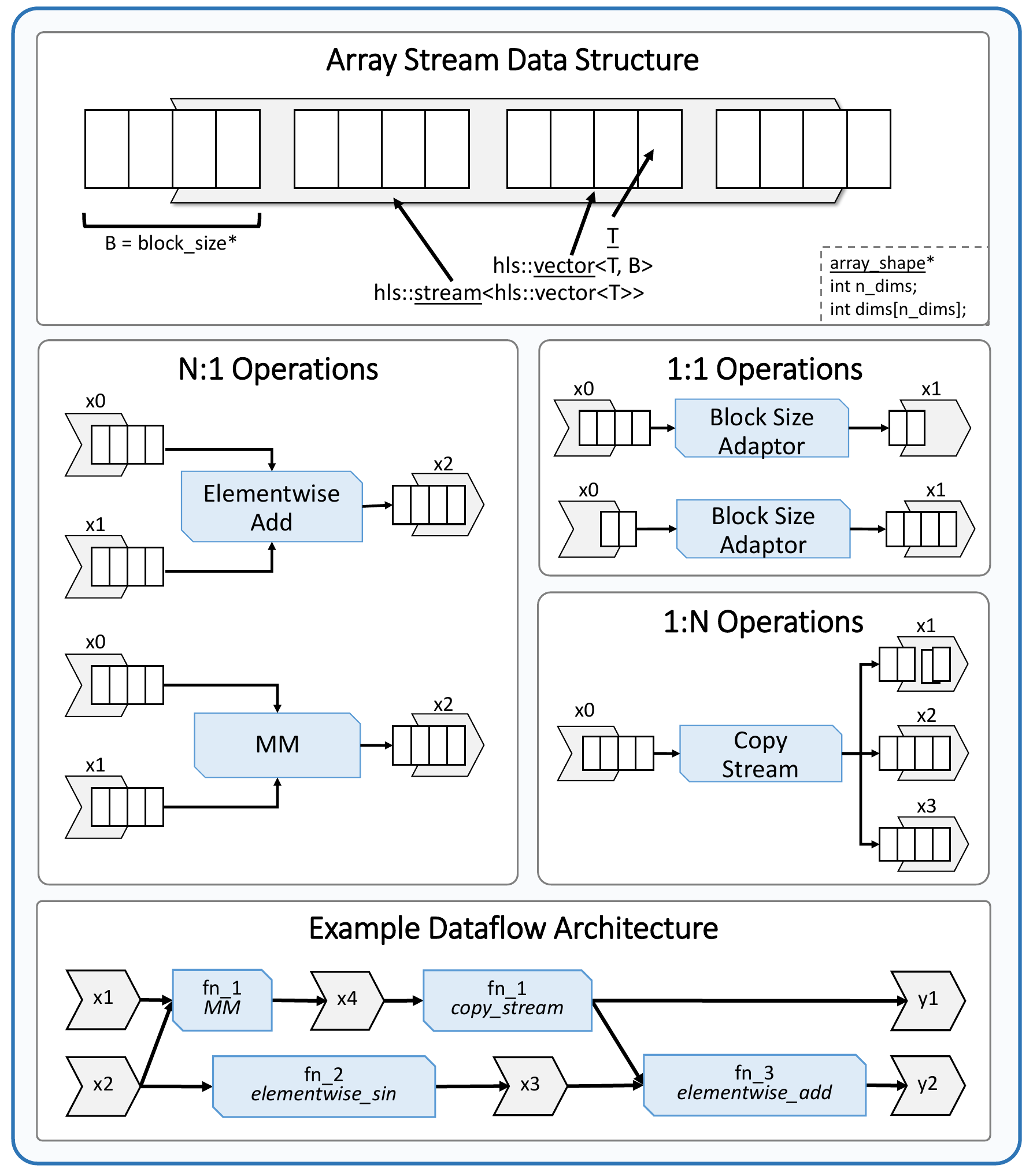}
    \caption{Illustration of the \texttt{array\_stream} data structure, the library of stream-based kernels, and an example compute graph mapped to a dataflow architecture.}
    \label{fig:dataflow_fig}
\end{figure}
\subsubsection{Challenges}

INR editing presents several significant challenges in translating the computation graph to efficient hardware, which we outline below:

\begin{itemize}[leftmargin=*]
    \item \textbf{Many Intermediate Results with Redundant Data Movement:}
    The conventional method of buffering intermediate results into scratchpad memory becomes infeasible due to the large computation graph size and the required batch size for effective INR usage. Allocating a buffer for each computation kernel in the graph could dramatically increase memory requirements. Specifically, INR model inference usually involves sampling multiple coordinates simultaneously to reconstruct data points, like pixel locations in an image. Consequently, the models demand an input batch size dimension, which propagates through all computation kernels. If the batch size is substantial, such as 64, it can inflate the memory requirement of scratchpad memories by 64 times.
    
    \item \textbf{Different Computation Kernels with Different Computation Patterns:}
    Inherent in the INR model is the use of diverse computation kernels, each exhibiting unique computation patterns. The diverse nature of these kernels means they may process data in distinct ways---some might favor sequential processing, while others may benefit from parallelization. This variance introduces a layer of complexity in optimizing the model as a whole and developing accelerated computation kernels. Effective use of these kernels requires careful coordination and optimal resource allocation to balance the computational load and data movement while minimizing area and latency requirements.
\end{itemize}

% We address the latter two points in the following section and leave the first to be addressed later in Sec.~\ref{sec:computation-graph}.

\subsubsection{Solution}

To address these challenges, we propose a \textbf{dataflow architecture} for mapping INR editing models to hardware. A visual overview is shown in Fig. \ref{fig:dataflow_fig}.

Our dataflow architecture is based on two key components: streaming-based data movement using a proposed \emph{array stream} data structure and a library of computational kernels designed to operate on array streams.

% \textbf{Streams address the issue of "Many Intermediate Results with Redundant Data Movement"}. Streams are modeled as FIFOs (both abstractly and in hardware) with a fixed size, also referred to later as "FIFO depth". We define our own variant of streams, called \emph{array streams}, which contain added metadata about the array shape, stream sizes, and block size of the data they represent, enabling the intermediate results to be streamed in a structured manner. Array streams also always stream the data in row-major order.

\textbf{Streams address the issue of ``Many Intermediate Results with Redundant Data Movement.''} They are conceptualized and physically implemented as fixed-size First-In-First-Out (FIFO) streams with a user-definable depth. Our unique variant, called ``array streams,'' includes additional metadata about array shape, stream sizes, and block size of represented data, thereby facilitating the structured streaming of intermediate results. Array streams are designed to stream data in row-major order.

This model is advantageous as it allows inputs, outputs, and intermediate activations to be stored as streams rather than relying on buffers such as scratchpad memory. The streams only need to store a fraction of the elements for any given input, output, or intermediate activation, resulting in a memory-efficient implementation compared to traditional buffered computations in CPUs and GPUs. The quantity of data that can be accommodated in the hardware is determined by the FIFO depth. Generally, we find that the FIFO depth can be significantly smaller than the total elements represented by the array stream, leading to substantial memory savings as outlined in Sec.~\ref{sec:fifo-depth-opt-results}.

Computational kernels, or compute units designed to interact with data streams, benefit from the array stream's unified interface. Each kernel is specialized to read and write data in its unique pattern. For instance, some kernels can instantly read and write computed data without buffering (e.g., elementwise add), while others may necessitate buffering (e.g., matrix multiply / MM) or access to array shape data (e.g., dimension select). Kernels are also categorized by their input-output degree: N:1, 1:1, and 1:N. \ourwork{} incorporates a subset of kernels necessary for supporting operations within INR-specific autograd computation graphs (refer to the source code for further exploration of all kernels).

When integrating array streams and computation kernels, the proposed dataflow architecture adheres to the ``one-producer, one-consumer'' principle. This necessitates that N:1 and 1:1 kernels be capable of mapping their outputs to inputs of downstream computation kernels while following the ``one-producer, one-consumer'' rule. To achieve this, a special 1:N operation known as ``copy stream'' is used to multicast a single input stream's elements to multiple output streams in a round-robin fashion.

The bottom panel in Fig. \ref{fig:dataflow_fig} shows an example of a mapped dataflow architecture for a small computation graph. In general, our work applies this dataflow architecture to larger extracted computation graphs, mapping inputs, outputs, and intermediate activations to array streams and operations to computation kernels.

\subsection{Compiler Methodology}

\subsubsection{Challenges}

The dataflow architecture provides a solid foundation for an efficient accelerator, but mapping a gradient computation graph onto this structure presents its own challenges:

\begin{itemize}[leftmargin=*]
    \item \textbf{Complex Computation Graphs with Redundant Operations:} In applying the INSP-Net approach, we build computation graphs by calculating higher-order gradients of the base INR model being edited. This process causes the computation graphs to grow exponentially with each gradient order. Both the base model and the higher-order composition graph share the same computations and redundant sub-graphs. Furthermore, there is an increase in redundant operations within these computation graphs, along with patterns of operations that can cancel each other out, leading to higher redundancy.
    \item \textbf{Susceptibility to Deadlock:} Due to the differing computation patterns of different computation kernels, the generated dataflow architecture is susceptible to deadlock unless FIFO buffers between kernels are carefully provisioned. It is critical to ensure that the generated design will not deadlock, but deadlocks are usually difficult to detect without a full cycle-level simulation, which can take hours or even days for the complex dataflow designs generated by our framework.
    \item \textbf{Latency or Memory Waste from Improper FIFO Buffer Sizing:} Deadlock-free operation is a necessary but not sufficient criterion for an efficient accelerator. Even when there is no deadlock, too-small FIFO depths can degrade performance so the resulting latency is multiple times slower than peak performance. On the other hand, too-large FIFO depths can consume multiple times the memory resources of an equally performant smaller design.
    \item \textbf{Complexity, Correctness, and Runtime Overhead of HLS Code:} Code generation can be an error-prone process. It is important to ensure that the generated code faithfully reproduces the gradient computation carried out by PyTorch while incurring minimal runtime overhead. The generated code should be as simple as possible to aid debugging and minimize the chance of errors.
\end{itemize}

We propose a four-step compilation process (represented in Fig.~\ref{fig:main_fig} as steps 2--5) to address each of these challenges.

\subsubsection{Computation Graph Extraction \& Optimization}
\label{sec:computation-graph}

\begin{figure}
    \centering
    \includegraphics[width=\linewidth]{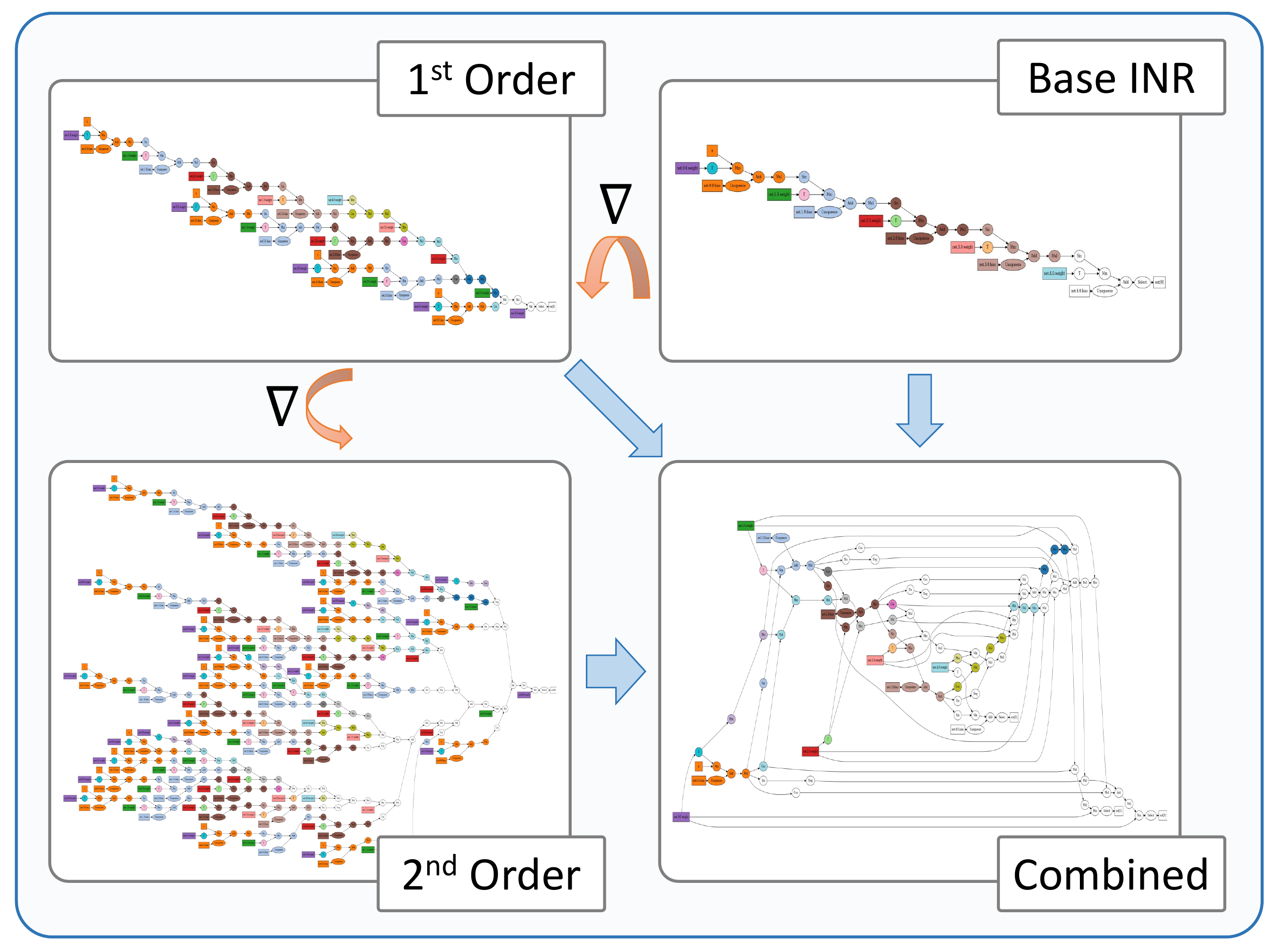}
    \caption{Visualization of the computation graph merging optimization. Similar computations are indicated with identical colors to represent their presence both within and across graphs. The merging of these graphs effectively minimizes redundant computations.}
    \label{fig:compute_graph_fig}
\end{figure}

The first step of our proposed process is to obtain the computation graph of the higher-order gradient of a desired function expressed as a series of PyTorch operations. We take advantage of the computation graph that is automatically built by PyTorch for its automatic differentiation process (``autograd'') as described in Sec.~\ref{sec:background-high-order-gradients}.

Given a list of PyTorch tensors representing the gradient outputs, we perform a depth-first traversal through the autograd graph of each of the tensors. We construct a combined computation graph from all the output tensors and apply several optimization passes to eliminate redundancy in the graph.

First, since the gradient introduces repeated subsections of the graph due to the chain rule of differentiation, we de-duplicate any common subtrees within the raw graph, indicated by the color-coded sections of Fig.~\ref{fig:compute_graph_fig}. As a result of this de-duplication, the output tensors across multiple gradient orders share most of their computation: for instance, the outputs for the \first{}-order gradient are contained entirely within the computation graph of the 2\textsuperscript{nd}-order gradient, with the exception of a few nodes at the end.

Second, the graph can contain ``Permute'' nodes, which perform an arbitrary permutation of the axes of the input tensor. However, in many cases, these ``Permute'' nodes simply swap the axes of a two-dimensional input, which is the same as transposing the input. Therefore, when we identify this special case anywhere in the graph, we replace the ``Permute'' node with a ``T'' (transpose) node.

Third, since transposing a tensor twice is the same as not modifying it at all, we look for any contiguous sequences of ``T'' nodes in the graph and remove all matched pairs, leaving zero or one ``T'' node in place of each sequence.

Finally, when multiple ``T'' nodes have the same input, we choose one of them to be the canonical node, delete the others, and re-route their outputs to come from the canonical node.

These optimizations massively simplify the graph. De-duplication greatly shrinks the graph size, making it feasible to synthesize accelerators for larger gradient computations. ``T'' node optimizations help reduce latency significantly, since transposing a tensor requires buffering the entire tensor and thus creates a bottleneck in the dataflow.

\subsubsection{Deadlock Analysis}
\label{sec:deadlock-analysis}

\begin{figure}
    \centering
    \includegraphics[width=0.9\linewidth]{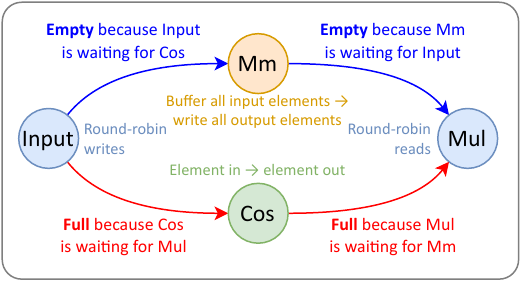}
    \caption{An example of a computation graph that causes deadlock with default FIFO sizing for any non-trivial input. The root cause is the contention between the ``Mm'' which buffers elements with a delay before writing out data and ``Cos'' which writes out data every cycle.}
    \label{fig:deadlock}
\end{figure}

Given an optimized computation graph, we must determine suitable buffer sizes for the FIFO streams connecting each kernel to avoid a deadlock in the overall design. To clarify how this issue arises, Fig~\ref{fig:deadlock} depicts an example computation graph that is susceptible to deadlock.

Two nodes, Mm and Cos, use the same input and feed the same output, but Cos operates in a fully streaming manner---producing each output element as soon as each input element is available---whereas Mm must fully buffer all the elements from this input before it can produce any output elements. The source node distributes outputs to Mm and Cos in a round-robin fashion, first writing one element to Mm, then the same element to Cos, repeating until all elements are written to both streams. Similarly, the Mul node reads input elements round-robin, reading one element from Mm, then one element from Cos, repeating until both streams are exhausted.

If all FIFOs use their default depth of 2 and there are more than five outputs from the source node, this computation graph is guaranteed to cause a deadlock:
\begin{enumerate}
    \item Mul will first attempt to read an element from the output of Mm.
    \item However, Mm will not produce an output until it reads all the elements from the source node.
    \item Meanwhile, Cos will attempt to write its outputs to Mul, which is blocked waiting for Mm's output; thus, after two output elements, the output stream for Cos will become full, blocking Cos from consuming more elements from its input stream.
    \item As a result, when the source node attempts to write the fifth output element to the input of Cos, it will stall, thus preventing Mm from receiving any more input.
\end{enumerate}
All four nodes in the computation graph become stalled waiting for each other cyclically, resulting in deadlock.

In this simple example, it is easy to see the cause of the deadlock and to determine a resolution: increase the stream depth of Cos's input to the total number of elements. However, the computation graphs for higher-order gradients can contain hundreds of nodes, thereby introducing complex dependency chains that cannot be analyzed by hand. Repeated simulation of the dataflow design with different FIFO depths is also infeasible, as the number of FIFOs involved in such a large computation graph leads to a massive design space. Thus we need a systematic approach to detecting and resolving deadlocks.

\begin{figure}
    \centering
    \includegraphics[width=1.0\linewidth]{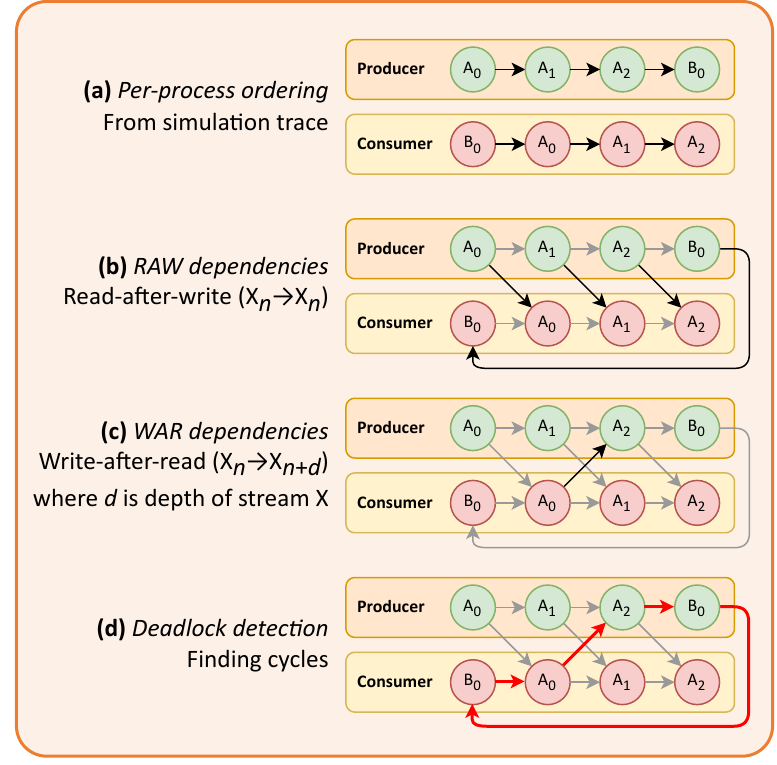}
    \caption{An example showing how the dataflow graph is constructed and used to detect deadlocks by searching for cycles. This example involves two FIFOs, A and B, both with depth 2. Green nodes represent FIFO writes; red nodes represent FIFO reads.}
    \label{fig:dfg}
\end{figure}
Our proposed solution is a \textit{dataflow graph} where nodes represent individual FIFO I/O operations (reads and writes) and directed edges represent ``happens-before'' relations. This graph encodes the entire behavior of a dataflow architecture with a given set of FIFO depths, and thus it can be used to determine precisely whether or not there will be a deadlock for some set of FIFO depths.

Fig.~\ref{fig:dfg} shows a simple example of the construction of this graph. In this example dataflow design, a producer process writes to two streams which are then read by a consumer process. Stream A transfers three data elements, A\textsubscript{0}, A\textsubscript{1}, and A\textsubscript{2}, and stream B transfers one, B\textsubscript{0}. Both streams have FIFO depth 2.

We start by determining the ordering of FIFO reads and writes within each process, as shown in Fig.~\ref{fig:dfg}(a). To obtain this ordering, we run our dataflow design through LightningSim~\cite{lightningsim}, a trace-based cycle-level simulator for HLS designs. The trace that LightningSim generates internally precisely orders all FIFO operations on a function-by-function basis. FIFO operations that must occur at the same time are grouped into one node, and edges connect nodes in the order defined by the trace. This trace only needs to be generated once for a given design, as the trace order is independent of the FIFO depths.

Then, in Fig.~\ref{fig:dfg}(b), we encode read-after-write (RAW) dependencies into the graph by adding edges connecting each write to its corresponding read: read \#$n$ from stream X cannot occur before write \#$n$ to stream X. This establishes the ordering of nodes between dataflow processes. As with Fig.~\ref{fig:dfg}(a), this is independent of FIFO depths and only needs to be done once for a given design. The resulting dataflow graph can be interpreted as the dataflow graph for a design where the FIFO depths are ``infinite'' or unconstrained.

Fig.~\ref{fig:dfg}(c) shows the encoding of write-after-read (WAR) dependencies into the graph. WAR dependencies are caused by limited FIFO depths: if a stream X has a depth of $d$, after $d$ writes to the stream, the stream will be full unless or until at least one read has occurred from the stream. Therefore, write \#$d$ depends on read \#$0$. Following similar logic, it follows that any write \#$n$ where $n \geq d$ depends on read \#$(n - d)$. In Fig.~\ref{fig:dfg}(c), with both FIFO depths set to 2, only write A\textsubscript{2} depends on read A\textsubscript{0}.

Finally, Fig.~\ref{fig:dfg}(d) demonstrates the deadlock detection algorithm, which is equivalent to finding cycles in the graph. Since edges represent ``happens-before'' relations, cycles represent that a node must happen before itself for the computation to proceed, which clearly represents a deadlock. In the figure, the write to A\textsubscript{2} must occur before the write to B\textsubscript{0} (by intra-process order), the write to B\textsubscript{0} must occur before the read from B\textsubscript{0} (by RAW dependency), the read from B\textsubscript{0} must occur before the read from A\textsubscript{0} (by intra-process order), and the read from A\textsubscript{0} must occur before the write to A\textsubscript{2} (by WAR dependency).

To resolve a deadlock, the depths of one or more of the streams with a WAR dependency in the cycle must be increased. In this example, the only WAR dependency in the cycle involves stream A, whose depth must be increased from 2 to 3 to resolve the deadlock.

Different combinations of stream depths can be quickly tested for deadlock by starting from the unconstrained graph, containing only intra-process and RAW dependencies, then adding WAR dependencies according to the stream depths and checking for cycles.

\subsubsection{FIFO Depth Optimization}
\label{sec:fifo-depth-opt}

Even if we determine a set of FIFO depths that are deadlock-free, it might be far from peak performance, or it might use excessive resources compared to similarly performant designs. We need a procedure to determine the peak performance of the design and find a set of FIFO depths that achieve similar performance without using excessive memory for FIFO buffers.

Luckily, the dataflow graph from Sec.~\ref{sec:deadlock-analysis} also allows us to estimate the latency of a dataflow design by assigning a minimum delay to each edge in the graph. We perform a topological sort on the nodes in the graph, then compute each node's latency as the maximum of its predecessors' latencies combined with the edge delays. The maximum latency across all nodes in the graph is a very close estimate to the latency of the overall design, excluding stalls incurred by, e.g., off-chip DRAM reads and writes.

Using these latency estimates, we are able to minimize memory usage without impacting performance. We start with the unconstrained graph and compute its latency estimate, which represents the peak performance of the design. Then, one by one, we constrain the depth of each stream to 2---the minimum depth for a FIFO queue---and re-run the latency estimator to see if the constraint changes the overall latency significantly (by more than a threshold $\alpha$, which is set to 1\% in our implementation). If it does, we discard the constraint; otherwise, we accept the new constraint. Once all streams have been evaluated, we run a simulation to determine the actual FIFO depths observed (peak number of FIFO queue slots used at any point in the simulation) under the newly added constraints. We use these observed numbers (with a minimum of 2 for each stream) as our final, optimized set of depths for all FIFOs in the computation graph.

\subsubsection{Code Generation}
\label{sec:code-gen}

The final HLS model is generated (and can be compiled and synthesized) using the code generation component of the presented framework. Code generation is done using a template-based compiler that maps kernels from the \ourwork{} hardware library to an HLS implementation of the model using the described dataflow architecture. Most of the implementation is simple initialization for \arrstream~ data structures and 1-to-1 mapping of functions in the computation graph to functions in the hardware library.

However, care needs to be taken when mapping hardware kernels to properly insert the hardware kernel calls in the correct topological order, as well as preserve the correct argument order from the computational graph. Each intermediate activation's argument order is stored in the associated edge as an edge feature in the processed computations graph, which is then referenced during code generation to generate kernel call argument lists. Care also must be taken to insert \cpystream{} kernels after function calls to effectively ``multicast'' kernel outputs to the correct downstream kernel inputs. This is done by extracting the edges to successors in the computation graph. These edges then become the edges to which the kernel output is multicast using the \cpystream{} kernel.

The metadata associated with \arrstreams{} is stored as compile-time information in the \arrstream{} struct implementation. The importance of this compile-time information becomes clear when computation kernels access array shape data through \texttt{typename} template arguments. This vital information at compile-time during High-Level Synthesis (HLS) can be skillfully utilized within the computation kernels for operations such as unrolling and pipelining of loops, which are dependent upon the array shape and block size specific to an individual \arrstream{}, as well as static asserts to check properties about the input arrays (e.g., array sizes for MM). For a more comprehensive overview of using  modern C++ features to implement these compile-time design features in HLS, we direct interested readers to \cite{lahti_leveraging_2023} as well as our source code.

The Python API for code generation takes in the processed computation graph (Sec. ~\ref{sec:computation-graph}) along with the computed FIFO depths from the deadlock analysis and FIFO depth optimization (Sec. ~\ref{sec:deadlock-analysis}, Sec. ~\ref{sec:fifo-depth-opt}). Additionally, the user is able to specify the target FPGA board along with desired fixed-point precision for the implemented HLS model which maps to the Vitis HLS arbitrary-precision fixed-point data structures. The code generation also handles the automated generation, compilation, and execution of a C++ testbench using the fixed-point model. This helps a user experiment and tweak the fixed-point precision used for the best resource usage vs. accuracy trade-off for model inference. Lastly, code generation handles the automated synthesis of the generated HLS model and extraction of synthesis report data for analysis.

\section{Results}

\begin{figure}
    \centering
    \includegraphics[width=1.0\linewidth]{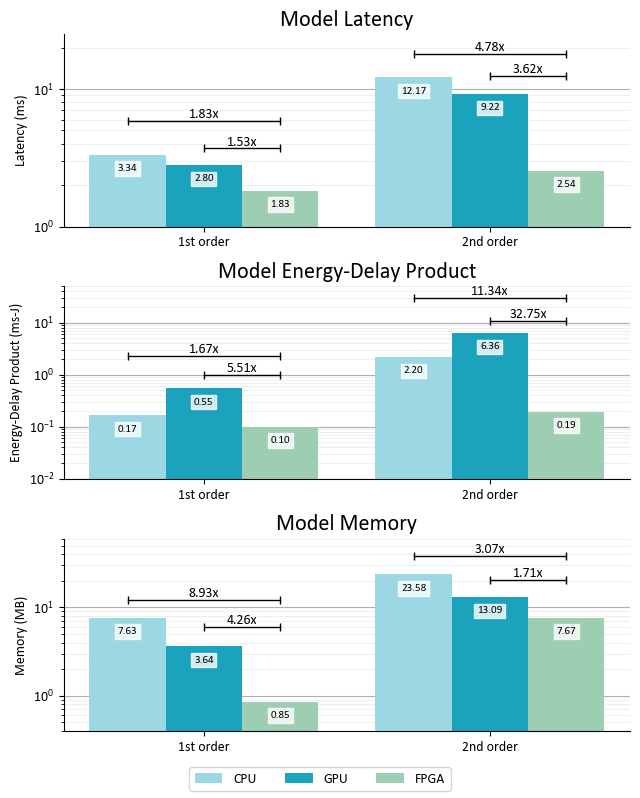}
    \caption{Main comparison results for latency, energy-delay product, and memory of 1st-order and 2nd-order INR models between GPU, CPU, and the proposed FPGA implementation. The y-axes are log scales.}
    \label{fig:results_bar}
\end{figure}

\subsection{Evaluation Setup}

As a case study to evaluate our framework, we measure the performance of two models derived from Xu \textit{et al.}~\cite{inrdsp}, a recent computer vision work that uses high-order gradients of a SIREN model~\cite{siren} to apply a variety of image transformations, such as blurring or denoising, directly to an image encoded as a SIREN INR.

We evaluate two configurations, namely, the first-order and second-order gradients of the SIREN model as computed in \cite{inrdsp}, using batch size 64 in both cases. The design for each of these two configurations was generated by the framework and synthesized using Xilinx Vitis HLS for the Xilinx Alveo U50 Data Center Accelerator at 300 MHz. A 32-bit fixed-point format with 10 integer bits was used for the data.

For the first-order model, a hardware parallelism factor of 64\texttimes{} was used for all MM operations. However, since the computation graph of the second-order model is so much more complex than that of the first-order model, the second-order model must use a lower parallelism factor of 16\texttimes{} for all MMs in order to avoid exceeding available resources on the target device.

FPGA latency results are collected using a highly accurate cycle-level simulator for HLS designs~\cite{lightningsim}, while resource estimates are provided by the HLS tool itself. Baseline results on CPU (Intel Xeon Gold 6226R) and GPU (NVIDIA RTX A6000) were measured directly from the gradient computation code in \cite{inrdsp}, written using the PyTorch framework.

% \subsection{Implementation Details}
\subsection{Latency, Energy Efficiency, and Memory}
\label{sec:main_results}

% Initial latency and energy efficiency results are presented for a first-order INR-DSP model, which involves computing the base MLP model output and its first-order gradient. The design was generated by the framework and synthesized using Xilinx's Vitis HLS for the Xilinx Alveo U50 Data Center Accelerator at 300 MHz. A hardware parallelism factor of 16$\times$ was used for all MM operations, and a 32-bit fixed-point format with 10 integer bits was used for the data.

% \input{floats/results_table}
\begin{table}%[h!] Inline tables are ugly, sorry -Rishov
\centering
% \footnotesize
% should we make it two column?
% i.e. spanning whole page?
% let me try
% feels too whitespacey
% \setlength{\tabcolsep}{2pt}
% yea try some different formatting, im always bad at latex tables
% lol no worries. I usually don't separate the absolute and relative columns
\caption{Performance comparisons.}
\begin{tabular}{cc|ccc}
\toprule
Model & Device & Latency (ms) & Memory (MiB) & EDP (J$\cdot$ms) \\
\midrule
1\textsuperscript{st} Order & CPU & 3.34 {\scriptsize(1.83\texttimes{})} & 7.63 {\scriptsize(8.93\texttimes{})} & 0.17 {\scriptsize(1.67\texttimes{})} \\
1\textsuperscript{st} Order & GPU & 2.80 {\scriptsize(1.53\texttimes{})} & 3.64 {\scriptsize(4.26\texttimes{})} & 0.55 {\scriptsize(5.51\texttimes{})} \\
1\textsuperscript{st} Order & FPGA & \textbf{1.83 {\scriptsize(1.00\texttimes{})}} & \textbf{0.85 {\scriptsize(1.00\texttimes{})}} & \textbf{0.10 {\scriptsize(1.00\texttimes{})}} \\
\midrule
2\textsuperscript{nd} Order & CPU & 12.17 {\scriptsize(4.78\texttimes{})} & 23.58 {\scriptsize(3.07\texttimes{})} & 2.20 {\scriptsize(11.34\texttimes{})} \\
2\textsuperscript{nd} Order & GPU & 9.22 {\scriptsize(3.62\texttimes{})} & 13.08 {\scriptsize(1.71\texttimes{})} & 6.36 {\scriptsize(32.75\texttimes{})} \\
2\textsuperscript{nd} Order & FPGA & \textbf{2.54 {\scriptsize(1.00\texttimes{})}} & \textbf{7.67 {\scriptsize(1.00\texttimes{})}} & \textbf{0.19 {\scriptsize(1.00\texttimes{})}} \\
\bottomrule
\end{tabular}

\vspace{2pt}
{\scriptsize Comparison factors (parenthesized) are relative to the corresponding FPGA metric.}
\label{tab:results_table_v2}
\end{table}

% also add somthing in the caption to explain the speedup numbers (x) are always realtive to FPGA idk if that is clear, but its that way to math the graphs

% \begin{table}[h]
% \centering
% \caption{Resource usage for our Base INR + 1st Order Gradient INR-DSP model.}
% \label{tab:resource_table}
% \begin{tabular}{lccccc}
% \toprule
% & \multicolumn{5}{c}{\textbf{Resources on the Alveo U50}} \\ \cmidrule(lr){2-6}
% & \textbf{BRAM\_18K} & \textbf{DSP} & \textbf{FF} & \textbf{LUT} & \textbf{URAM} \\ 
% \midrule
% Available             & 2688      & 5952& 1743360& 871680  & 640  \\
% Utilization (\%)      & 12        & 15  & 10      & 36      & 0    \\
% \bottomrule
% \end{tabular}
% \end{table}

\begin{table}
\centering
\caption{Resource usage vs. latency on the Alveo U50.}
% for our Base INR + \first{} Order Gradient INR-DSP model.}
\setlength{\tabcolsep}{5pt}
% \renewcommand{\arraystretch}{0.8}
% \begin{tabular}{lcc|ccccc}
% % \cmidrule(lr){2-6}
% % & \multicolumn{5}{c}{\textbf{Alveo U50 Resources}} \\ \cmidrule(lr){2-6}
% \toprule
% Model & MM Parallel & Latency (ms) & BRAM & DSP & FF & LUT & URAM \\
% \midrule
% \midrule
% Available             &&& 2688      & 5952& 1743360& 871680  & 640  \\
% % Total                 & 347       & 895 & 186803  & 320992  & 0    \\
% % \textbf{Utilization}      & \textbf{12\%}        & \textbf{15\%}  & \textbf{10\%}      & \textbf{36\%}      & \textbf{0\%}   \\
% \bottomrule
% \end{tabular}
\begin{tabular}{r|ccc|c}
\toprule
Model & \first{} Order & \first{} Order & 2\textsuperscript{nd} Order \\
MM Parallelism & 64\texttimes{} & 16\texttimes{} & 16\texttimes{} \\
\cmidrule{1-4}
Latency (ms) & 1.83 & 2.55 & 2.54 & Available \\
\midrule
BRAM & 389 {\scriptsize (14\%)} & 233 {\scriptsize (9\%)} & 419 {\scriptsize (16\%)} & 2,688 \\
DSP & 3,343 {\scriptsize (56\%)} & 1,039 {\scriptsize (17\%)} & 3,889 {\scriptsize (65\%)}& 5,952 \\
FF & 529k {\scriptsize (30\%)} & 277k {\scriptsize (16\%)} & 952k {\scriptsize (55\%)} & 1,743k \\
LUT & 367k {\scriptsize (42\%)} & 234k {\scriptsize (27\%)} & 781k {\scriptsize (90\%)} & 871k \\
URAM & --- & 48 {\scriptsize (8\%)} & 192 {\scriptsize (30\%)} & 640 \\
\bottomrule
\end{tabular}
\label{tab:resource_table}
\end{table}

Results are shown in Table~\ref{tab:results_table_v2} and Table~\ref{tab:resource_table}. Across both first-order and second-order models, the FPGA implementation \textbf{beats CPU and GPU baselines in three key metrics: latency, memory usage, and energy-delay product.}

Our first-order gradient model on FPGA achieves a significant speedup over CPU and GPU baselines, but the speedup achieved by our second-order gradient model is even more pronounced, where the generated accelerator achieves nearly 4\texttimes{} speedup over GPU and nearly 5\texttimes{} over CPU.

Notably, Table~\ref{tab:resource_table} shows how, when the same MM parallelism factor is used for different-order gradients, the latencies of the resulting accelerators are very similar. This demonstrates the advantage of our dataflow architecture: because we can overlap most of the kernels in the computation graph, a larger computation graph induced by a higher-order gradient does not always mean the latency will be significantly higher. Even when MM parallelism must be reduced for the model to fit within the target device's resources, it does not result in an increase in latency by the same factor.

(That the 1\textsuperscript{st}-order model with 16\texttimes{} MM parallelism is slightly slower than 2\textsuperscript{nd}-order model with 16\texttimes{} MM parallelism may initially appear erroneous, given that the 1\textsuperscript{st}-order computation graph is a subset of the 2\textsuperscript{nd}-order graph. However, it is a result of our FIFO depth optimization process and will be explained in Sec.~\ref{sec:fifo-depth-opt-results}.)

We also see significant memory savings over CPU and GPU baselines, about 9\texttimes{} less memory than CPU and 4\texttimes{} less memory than GPU on the 1\textsuperscript{st}-order model and about 3\texttimes{} and 2\texttimes{} less than CPU and GPU on the 2\textsuperscript{nd}-order model.

Our framework demonstrates its strongest advantage in energy efficiency over CPU and GPU baselines: our model achieves an energy-delay product over 11\texttimes{} lower than CPU and nearly 33\texttimes{} lower than over GPU on the 2\textsuperscript{nd}-order model, thanks to the combination of low latency and low power achieved by our FPGA design.
% A \textbf{1.72$\times$ speedup over CPU}, a \textbf{1.5$\times$ speedup over GPU}, and a \textbf{3.66$\times$ increase in energy efficiency compared to GPU} are demonstrated. It is also shown that the current model using first-order gradients can easily fit onboard, leaving room for ongoing work to extend the models to computation graphs for higher orders.

\subsection{Graph Optimization}

\begin{table}
\centering
\caption{Computation graph optimizations.}
\setlength{\tabcolsep}{1.65pt}
\begin{tabular}{l|cc|ccc}
\toprule
&&& \multicolumn{3}{c}{Node Types} \\
Optimization & Nodes & Edges & T & Permute & Other \\
\midrule
Original graph & 5,531 & 7,326 & 438 & 945 & 4,148 \\
+ Dedupe common subtrees & 459 {\scriptsize (\textminus 92\%)} & 626 {\scriptsize (\textminus 91\%)} & 63 & 5 & 391 \\
+ Replace ``Permute''s \textrightarrow{} ``T''s & 459 {\scriptsize (\textpm 0\%)} & 626 {\scriptsize (\textpm 0\%)} & 68 & 0 & 391 \\
+ Remove ``T'' pairs & 420 {\scriptsize (\textminus 8\%)} & 587 {\scriptsize (\textminus 6\%)} & 29 & 0 & 391 \\
+ Dedupe common ``T''s & 396 {\scriptsize (\textminus 6\%)} & 563 {\scriptsize (\textminus 4\%)} & 5 & 0 & 391 \\
\bottomrule
\end{tabular}
\label{tab:graph_opt_table}
\end{table}

We perform an ablation study of our computation graph optimization techniques described in Sec.~\ref{sec:computation-graph} and report our findings in Table~\ref{tab:graph_opt_table}. The most significant optimization is the de-duplication of common subtrees in the graph, which accounts for over 90\% reduction in both nodes and edges over the unoptimized graph. However, the other optimizations we perform result in significant drops in the number of ``Permute'' and ``T'' nodes, collectively dropping their combined total from 68 nodes to just 5. This minimizes bottlenecks in the dataflow computation, as ``Permute'' and ``T'' both require buffering the entire input stream before writing outputs.

\subsection{FIFO Depth Optimization}
\label{sec:fifo-depth-opt-results}

\begin{table}
\centering
\caption{Before and after FIFO depth optimization.}
\setlength{\tabcolsep}{3pt}
\begin{tabular}{cc|cc|cc}
\toprule
&& \multicolumn{2}{c|}{Before Optimization} & \multicolumn{2}{c}{After Optimization} \\
Model & MM\,$\parallel$ & Latency (ms) & $\sum$\,Depths & Latency (ms) & $\sum$\,Depths \\
\midrule
1\textsuperscript{st} Order & 64\texttimes & 1.823 & 125,586 & {\renewcommand{\arraystretch}{0.9}\begin{tabular}[t]{@{}c@{}}1.828\\{\scriptsize (+0.3\%)}\end{tabular}} & {\renewcommand{\arraystretch}{0.9}\begin{tabular}[t]{@{}c@{}}15,579\\{\scriptsize (\textminus 87.6\%)}\end{tabular}} \\
1\textsuperscript{st} Order & 16\texttimes & 2.538 & 125,661 & {\renewcommand{\arraystretch}{0.9}\begin{tabular}[t]{@{}c@{}}2.551\\{\scriptsize (+0.5\%)}\end{tabular}} & {\renewcommand{\arraystretch}{0.9}\begin{tabular}[t]{@{}c@{}}15,643\\{\scriptsize (\textminus 87.6\%)}\end{tabular}} \\
2\textsuperscript{nd} Order & 16\texttimes & 2.545 & 668,601 & {\renewcommand{\arraystretch}{0.9}\begin{tabular}[t]{@{}c@{}}2.545\\{\scriptsize (+0.0\%)}\end{tabular}} & {\renewcommand{\arraystretch}{0.9}\begin{tabular}[t]{@{}c@{}}96,808\\{\scriptsize (\textminus 85.5\%)}\end{tabular}} \\
\bottomrule
\end{tabular}

\vspace{2pt}
{\scriptsize MM\,$\parallel$ = MM parallelism; $\sum$\,Depths = Sum of FIFO depths}
\label{tab:fifo_depth_table}
\end{table}

We also evaluate the effectiveness of the FIFO depth optimization scheme described in Sec.~\ref{sec:fifo-depth-opt} in reducing memory usage. We consider two metrics: the latency of the model and the sum of FIFO depths, which acts as a proxy for the memory consumed by the FIFOs. We evaluate each metric both before and after optimization, where the set of FIFO depths before optimization is determined as the depths actually observed (with a minimum of 2 for each stream) when we run a simulation with all FIFO depths unconstrained (i.e., a simulation of peak performance).

Table~\ref{tab:fifo_depth_table} shows our results. In all three cases evaluated, we achieve over 85\% reduction in FIFO depths with less than 1\% degradation over peak performance.

These results also explain why the 1\textsuperscript{st}-order model with 16\texttimes{} MM parallelism runs slightly slower than the 2\textsuperscript{nd}-order model with 16\texttimes{} MM parallelism, despite the 1\textsuperscript{st}-order graph being a subset of the 2\textsuperscript{nd}-order graph. At peak performance, the 1\textsuperscript{st}-order model is slightly faster; however, the FIFO depths selected for these two models by the optimization process in Sec.~\ref{sec:fifo-depth-opt} end up causing the final latency of the 1\textsuperscript{st}-order model to slightly exceed the final latency of the 2\textsuperscript{nd}-order model. This can be avoided by adjusting the acceptable threshold $\alpha$ during depth optimization.

\subsection{Dataflow Trace Visualization}
\label{sec:dataflow-trace-viz}

Novel simulation tools \cite{lightningsim} are used to dump and inspect simulation traces to analyze FIFO read and FIFO writes and better understand data movement along \arrstreams.

\begin{figure}
    \centering
    \includegraphics[width=1.0\linewidth]{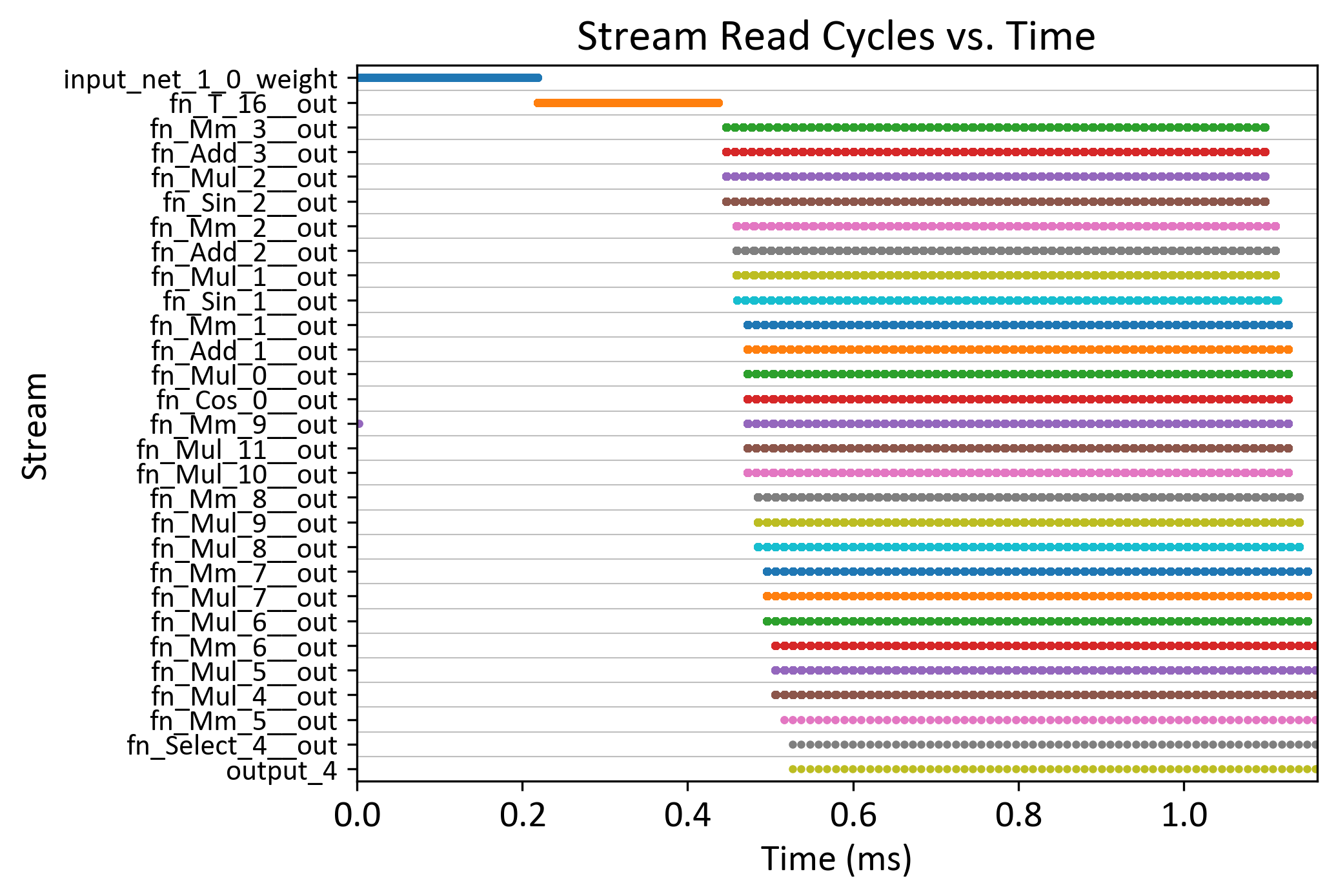}
    \caption{A trace of FIFO reads for a representative subset of hardware computation kernels in the main dataflow region of the Base INR + \first{} Order Gradient INR-DSP model.}
    % \caption{The read cycles of a representative subset of FIFOs in the generated design, which depict how data flows through the design.}
    \label{fig:trace}
\end{figure}

The FIFO reads over time during computationally intensive operations, mainly matrix multiplication (MM), are shown in Fig. \ref{fig:trace}. Due to the ordering of dependencies in the computation graph, it is clear when some MM operations are computing in parallel, as well as when data is being stalled periodically for downstream computation kernels. Work is ongoing to show other complex simulation behavior of the dataflow to better understand FIFO depths over time for better FIFO sizing and deadlock detection.

\section{Conclusion}

In this paper, we introduced INR-Arch, a framework for dataflow architectures of \nth-order gradient computations. This addresses the challenges that traditional architectures encounter when computing higher-order gradients efficiently. We centered our evaluation application on INR editing and compared our framework against CPU and GPU baselines. We demonstrated significant speed improvements, decreased memory usage, and a lower energy-delay product than both the CPU and GPU baselines.

Future work involves extending our evaluation to include higher-order gradients, examining the applicability of our framework to diverse models, and addressing large, intricate designs like those found in high-performance computing (HPC). These complex designs involve computational kernels or FIFO buffers that may not fit on the board. Furthermore, we plan to continue developing highly optimized and compact model caricatures for additional edge computing applications. By expanding our framework to handle higher-order gradients, we can further illustrate its adaptability and effectiveness across a wider range of applications. Moreover, our goal is to adapt our framework to suit different models, empowering researchers to utilize FPGA acceleration for a multitude of computational tasks beyond the INR editing scenario.

By providing an open-source implementation on GitHub, we invite further exploration, collaboration, customization, and deployment of our framework. This approach can serve the distinct needs of various research domains.

\section{Acknowledgements}

This work and its authors are partially supported by the Center for Research
into Novel Computing Hierarchies (CRNCH) at Georgia Tech, the 2022 Qualcomm Innovation Fellowship program, Cisco, and Georgia Tech Research Institute.

\bibliographystyle{IEEEtran}
\bibliography{refs}

\end{document}